\documentclass[aps,pre,twocolumn,groupedaddress,10pt,nofootinbib,a4paper,showpacs]{revtex4-1}

\usepackage{dcolumn}% Align table columns on decimal point
\usepackage{bm}% bold math
\usepackage{amsmath}
\usepackage{amssymb}
\usepackage{amsfonts}
\usepackage{amsthm}
\usepackage{amscd}
\usepackage{graphicx}
\usepackage{subfigure}
\usepackage{bm}

\def\be{\begin{equation}}
\def\ee{\end{equation}}

\def\bra#1{\mathinner{\langle{#1}|}}
\def\ket#1{\mathinner{|{#1}\rangle}}
\def\braket#1{\mathinner{\langle{#1}\rangle}}

{\catcode`\|=\active
  \gdef\set#1{\mathinner{\lbrace\,{\mathcode`\|"8000\let|\midvert #1}\,\rbrace}}
  \gdef\Set#1{\left\{\:{\mathcode`\|"8000\let|\SetVert #1}\:\right\}}}
\def\midvert{\egroup\mid\bgroup}
\def\SetVert{\egroup\;\mid@vertical\;\bgroup}

\newcommand{\mbf}[1]{\mathbf{#1}}

\begin{document}
\title{Transport efficiency in topologically disordered networks with environmentally induced diffusion}

\author{P. Schijven, J. Kohlberger, A. Blumen and O. M\"ulken}
\affiliation{Physikalisches Institut, Universit\"at Freiburg, Hermann-Herder-Strasse 3, 79104 Freiburg, Germany}

\date{\today}

\begin{abstract}
 We study transport in topologically disordered networks that are
 subjected to an environment that induces classical diffusion. The dynamics is 
 phenomenologically described within the framework of the recently introduced 
 quantum stochastic walk, allowing to study the crossover between coherent 
 transport and purely classical diffusion. We find that the coupling to the environment removes 
 all effects of localisation and quickly leads to classical transport. Furthermore, we
 find that on the level of the transport efficiency, the system can be well described
 by reducing it to a two-node network (a dimer).
\end{abstract}


\maketitle

\section{Introduction}
Recent experimental advances in ultra-cold Rydberg gases
allow for a precise control of its constituent atoms, including
the strength of their interactions by the use of F\"orster resonances \cite{Anderson1998}.
At these temperatures the thermal energy is much smaller than the interaction 
energies and the interaction dynamics is much faster than the 
displacement due to thermal motion. The dynamics is therefore
completely driven by Rydberg-Rydberg interactions \cite{Westermann2006}. In experiments,
one usually excites one atom into a Rydberg state that is resonantly
coupled to a lower state. Resonant energy transfer now leads to 
quick coherent hopping of the excitation to nearby atoms \cite{Saffman2010}. Since the
large dipole moment of this Rydberg state leads to the dipole-blockade mechanism preventing 
the appearance of multiple Rydberg excitations in a nearby area \cite{Jaksch2000}, one can study
excitonic transport in this system by considering a network of 
coupled two-level systems \cite{Cote2006,Saffman2010}. 

Another important example of resonant energy transfer is the study of exciton
transport \cite{Kenkre1982} in light harvesting systems found in, for example, marine algae \cite{Reineker2004,Collini2010,Fleming2004}. 
After the absorption of a photon, the exciton is transfered along a series of
bacterio-chlorophylls (BChl) to the reaction center. Recent experiments indicate that this 
transport shows coherent features, even at room temperatures %, leading to highly efficient dynamics 
\cite{Cheng2006,Engel2007,Panitchayangkoon2010}. 

For both of these systems it is important to study the effects of environmental
interactions on the coherent transport of excitations \cite{Vaziri2010}. These interactions will
in general lead to decoherence and are therefore usually detrimental to the
transport efficiency. There are situations however, where the presence of 
onsite energetic disorder and decoherence rates that are proportional to 
the intra-site couplings may actually lead to faster transport \cite{Cao2009}. This is for
example demonstrated in \cite{Caruso2009} for a fully connected network and in \cite{Rebentrost2009, Mohseni2008}
for the FMO-complex found in light harvesting systems. 
In these works the environment is assumed to be weakly coupled to the system, leading to
Markovian transport. 

On the other hand, one can also take the following complementary viewpoint: instead of
modelling the exact type of environmental interactions by phenomenological
master equations, one can also design a system-bath coupling to engineer systems that
exhibit efficient transport. This idea was for example suggested in \cite{Diehl2008}
and \cite{Verstraete2009} where such a procedure for the use of quantum state engineering
has been proposed. 

Motivated by these viewpoints we consider systems for which the environmental interactions
induce incoherent transfer of the populations. These interactions have the advantage that
for large couplings to the environment the system does not run into the quantum Zeno effect,
but instead behaves as if it is governed by classical diffusion. By suitably engineering
the system and the coupling to the environment, we can therefore study the transition from
purely coherent dynamics, which is usually described by the continuous-time quantum walk \cite{Farhi1998,Muelken2011}, to
purely incoherent dynamics that is described by classical diffusion (or the continuous-time
random walk) \cite{VanKampen1990}. 

For the mathematical framework underlying this transition, we use the recently introduced
quantum stochastic walk \cite{Whitfield2010}. This is a generalisation of the continuous-time quantum walk by
also allowing for incoherent transfer between the sites of the system. This model also shows
similarities to early approaches like the Haken-Strobl-Reineker model \cite{HakenStroblBook,Haken1972}. To study the transport
efficiency we furthermore connect our system with a source and a drain and model the transitions
from and to the system by irreversible incoherent transfer. This idea was already implemented by 
several groups \cite{Michaelis2006,Groth2006,Caruso2009,Vaziri2010,Sarovar2011}. 

In this paper we take our model system to be a topologically disordered network with long-range dipole-dipole
interactions, resembling for instance a gas of ultra-cold Rydberg atoms. We study the crossover between quantum and classical
transport and the transport behaviour as a function of the source and drain strengths. We find that
on the level of the transport efficiency, we can effectively reduce our network to a dimer.

\section{The quantum stochastic walk}
In this section we review the mathematical details of both the continuous-time random walk 
(CTRW) and the continuous-time quantum walk (CTQW) and we introduce the concept of the
quantum stochastic walk (QSW) that describes the transition between the CTRW and the CTQW.

\subsubsection{Random walk}
Consider a network consisting of $N$ nodes. One can specify the connections in
the network by the connectivity matrix $\mbf{A}$. This $N\times N$ matrix is defined by:
\be
  A_{kj} = \left\{ \begin{array}{cl}
                    f_j & \text{for $k=j$} \\
                    -1 & \text{if $k$ and $j$ are connected} \\
                    0  & \text{else}
                   \end{array}
           \right. ,
\ee
where $f_j$ is the functionality of the node $j$, i.e. the number of  bonds emanating from $j$.
To each node $k$ a vector $\ket{k}$ is associated such that the collection of 
all these vectors forms an orthonormal basis of $N$-dimensional vector space.

The dynamics on the network is governed by the transfer matrix $\mbf{T}$, which is
the matrix of transition rates per unit time. For the simplest case, where all
the transition rates are equal to, say $\gamma$, $\mbf{T}$ is related to the connectivity
matrix $\mbf{A}$ by $\mbf{T} = - \gamma \mbf{A}$. After assuming that the transport is described
by a Markovian process, one arrives at the master equation \cite{VanKampen1990}
\be\label{eq:mastereq-ctrw}
  \frac{d p_{kl}(t)}{dt} = \sum_{j=1}^N T_{kj} p_{jl}(t) ,
\ee
where $p_{kl}(t)$ is the probability of being at node $k$ after a time $t$, under the constraint
that one starts at $t=0$ in some node $l$. This master
equation is the defining evolution equation for a CTRW.

\subsubsection{Quantum walk}
The CTQW can be formulated in a similar fashion. One can
interpret the basis vectors $\ket{k}$ as forming a basis for the whole accessible Hilbert space. 
The main idea is now to identify the CTQW Hamiltonian $\mbf{H}$ with the CTRW transfer matrix $\mbf{T}$ 
by setting $\mathbf{H} = - \mbf{T}$  \cite{Farhi1998}. The dynamics of the system is then described by 
the Schr\"odinger equation
\be
 \frac{d}{dt} \alpha_{kl}(t) = - i \sum_j H_{kj} \alpha_{j l}(t),
\ee
for the transition amplitudes 
\be
  \alpha_{kl}(t) = \braket{k | e^{-i \mbf{H} t} | l} ,
\ee
which describe the overlap of the initial state $\ket{l}$ with the final state $\ket{k}$ after a time $t$.
Equivalently, one can also formulate the CTQW by specifying the evolution of the density
operator $\bm{\rho}(t)$, which is described by the Liouville-von Neumann equation:
\be
  \frac{d \bm\rho(t)}{dt} = -i \left[ \mbf{H}, \bm \rho(t) \right] .
\ee

\subsubsection{Quantum stochastic walk}
Now, if one places the
system in an external environment, the full Hamiltonian takes the form $\mbf{H}_{\text{tot}} = \mbf{H} + \mbf{H}_E + \mbf{H}_{\text{int}}$,
where $\mbf{H}$ is the Hamiltonian of the network, $\mbf{H}_E$ is the Hamiltonian of the environment
and $\mbf{H}_{\text{int}}$ specifies the interactions between the network and the environment. When the environmental
correlation time is small compared to the relaxation time of the system, one
can employ the Born-Markov approximation. This approximation results in the following general
form for the evolution equation of the reduced density operator $\bm{\rho}(t)$ of the system \cite{BreuerOpenQS, Lindblad1976},
which is called the \emph{Lindblad equation}:
\be
   \frac{d \bm{\rho}(t)}{dt} = -i \left[ \mbf{H}, \bm{\rho}(t) \right] + \sum_{k,l=1}^{N} \lambda_{kl} \mathcal{D}(\mbf{L}_{kl}, \bm{\rho}(t)) ,
\ee
with the constants $\lambda_{kl}\geq 0$ for all $k$ and $l$ and
\be
\mathcal{D}(\mbf{L}_{kl}, \bm{\rho}(t)) = \mbf{L}_{kl}^{\phantom{\dagger}} \bm{\rho}(t) \mbf{L}_{kl}^\dagger - 
\frac{1}{2}\left\{\mbf{L}_{kl}^\dagger \mbf{L}_{kl}^{\phantom{\dagger}}, \bm{\rho}(t) \right\} .
\ee
The operators $\mbf{L}_{kl}$ are called Lindblad operators and they form an orthonormal
basis for the space of operators acting on the system's Hilbert space. The constants $\lambda_{kl}$ are related to
certain correlation functions of the environment and they play the role of relaxation rates for
different quantum channels. For notational clarity we introduce the \emph{dissipator} $\mathcal{D}(\bm{\rho}(t))$ as
\be
  \mathcal{D}(\bm{\rho}(t)) = \sum_{k,l=1}^{N} \lambda_{kl} \mathcal{D}(\mbf{L}_{kl}, \bm{\rho}(t)) .
\ee
When it is clear from the context, the explicit dependence on $\bm{\rho}(t)$ will be omitted. 

The Lindblad equation is the defining equation for the quantum stochastic walk (QSW) \cite{Whitfield2010}. It 
contains both the coherent evolution due to the Hamiltonian and 
the incoherent evolution due to the environmental interactions. Until now we have not made any assumptions on the 
particular details of the coupling to the environment. Here we focus on the transitions between the CTQW and the CTRW, 
thus we consider environmental interactions that will eventually lead to classical diffusion. This can be achieved by a proper choice of
the Lindblad operators $\mbf{L}_{kl}$, as is shown in appendix \ref{app:randomwalk}. The dissipator corresponding to
these Lindblad operators describes a CTRW for the populations (the diagonal elements of $\bm{\rho}(t)$) and pure decoherence
leading to exponentially decaying coherences (the offdiagonal elements of $\bm{\rho}(t)$). 
Furthermore, we choose the coupling constants $\lambda_{kl}$ to be equal to the absolute value of the corresponding matrix elements
of the Hamiltonian, that is $\lambda_{kl} =  | H_{kl} | = |T_{kl}|$. This ensures that the CTRW has the same transfer rates as the
CTQW. 

The full evolution equation of our system is then given by a linear combination of the CTQW and the CTRW combined with pure decoherence,
together with a scaling parameter $\alpha$ \cite{Whitfield2010}:
\be\label{eq:qsw-alpha}
   \frac{d\bm{\rho}(t)}{dt} = (1-\alpha)\mathcal{L}_{\text{coh}}(\bm{\rho}(t)) + \alpha \mathcal{D}(\bm{\rho}(t)), \quad \alpha\in [0,1] ,
\ee
with $\mathcal{L}_{\text{coh}}(\bm{\rho}(t)) = -i \left[ \mbf{H}_S, \bm{\rho}(t) \right] $ representing the generator
for purely coherent transport. In the limit  $\alpha \to 0$ we obtain the CTQW and in the limit $\alpha\to 1$ we obtain the CTRW. 

\section{Sources and drains}\label{sec:sourcedrain}
\subsection{General considerations}
In order to incorporate sources and drains into the system, we consider
the following scenario: for a single source we augment the network of $N$ nodes by
an additional node $\ket{0} \equiv \ket{\text{source}}$. In order to prevent the excitation
from flowing back into the source this node will be
incoherently coupled to $S$ nodes of the system, with $1\leq S \leq N$. The incoherent
nature of the coupling implies that we do not couple the source to the network by
the Hamiltonian, but that we use the Lindblad formalism to describe an incoherent
hopping from the source to the network, i.e. $\mbf{H}_{kl}=0$ for
all the nodes $\ket{k}$ comprising the original network. 

Similarly, we include a drain as an extra node $\ket{N+1}\equiv\ket{\text{drain}}$ that is incoherently coupled
to $D$ nodes of the system, with $1\leq D \leq N$. Thus
the total dimension of the systems Hilbert space will be $N+2$ and 
the reduced density matrix of the source-network-drain system will be a 
$(N+2)\times(N+2)$ matrix.

When the source (drain) is coupled to more than one node of the network, there are
different ways to model the transition from the source (drain) to these nodes and
the resulting dynamics can vary strongly. However, in this paper we only consider
the simplest case where both the source and drain are connected to only one node,
leaving the more general case for future work \cite{SchijvenUnpublished}

The source node $\ket{0}$ will be coupled to node $\ket{k}$ of the network by introducing
the Lindblad operator $\mbf{L}_{k,0} = \ket{k}\bra{0}$ (see e.g. \cite{Michaelis2006, Groth2006}). 
We will denote its coupling strength by $\Gamma_{k}$. Similarly, the drain node $\ket{N+1}$ will be 
coupled to node $\ket{l}$ by introducing the Lindblad operator $\mbf{L}_{N+1, l} = \ket{N+1}\bra{l}$ 
and its respective coupling strength will be denoted by $\gamma_{l}$.

We pause to note that there is a difference between the $\Gamma_k$ and $\gamma_l$ introduced here and in
previous work \cite{Agliari2010,Mulken2007}.
% Care must be taken however in defining these coupling strengths $\Gamma_k$ and $\gamma_l$, since there is a 
% difference in definition between this approach and the approach that was used in previous work \cite{Agliari2010,Mulken2007}. 
There a drain connected to a node $\ket{m}$ was modelled by considering the effective Hamiltonian 
$\mbf{H}_{eff} = \mbf{H} - i \gamma_m \ket{m}\bra{m}$. As is shown in appendix \ref{app:effectiveHamilt-trap}, 
in the Lindblad approach this is equivalent to a coupling strength equal to $2\gamma_m$.

The dynamics of the reduced density operator $\bm{\rho}(t)$ is then given by the usual Lindblad equation for the quantum
stochastic walk plus the extra dissipators corresponding to the Lindblad operators that model the coupling of the source and drain
to the network:
\be\label{eq:lindbladplussource-drain}
   \frac{d\bm{\rho}(t)}{dt} = (1-\alpha)\mathcal{L}_{\text{coh}}(\bm{\rho}(t)) + \alpha \mathcal{D}(\bm{\rho}(t)) + \mathcal{L}_{s+d}(\bm{\rho}(t)) ,
\ee
with
\begin{eqnarray}\label{eq:generator-sourcedrain}
\mathcal{L}_{s+d} (\bm{\rho}(t)) &=& \Gamma_k \mathcal{D}(\mbf{L}_{k,0}, \bm{\rho}(t)) + \nonumber \\
                                && \gamma_l \mathcal{D}(\mbf{L}_{N+1,l}\, ,\bm{\rho}(t)) .
\end{eqnarray}
Note that both $\mathcal{L}_{\text{coh}}$ and $\mathcal{D}(\bm{\rho}(t))$ only act on the subspace spanned by the network nodes, 
while $\mathcal{L}_{s+d}$ acts on the combined source-network-drain system. 

As an initial condition we will always choose to start in the source node, so $\bm{\rho}(0) = \ket{0}\bra{0}$. By the explicit
form of the dissipators given by Eq. \eqref{eq:explicit-dissipators} and by using that $H_{0k} = H_{N+1,k} = 0$ for $k=0,\ldots N+1$,
we have the following expressions for the coherences between the source resp. drain and
the rest of the network:
\begin{eqnarray}
  \frac{d\rho_{j0}(t)}{dt} &=& -\frac{1}{2} \Gamma_k \rho_{j0}(t), \quad \forall j\neq 0 \label{eq:vanishing-coherences-1} \\
  \frac{d\rho_{j, N+1}(t)}{dt} &=& - \frac{1}{2} \gamma_l \rho_{j, N+1}(t), \quad \forall j\neq N+1 \label{eq:vanishing-coherences-2}.
\end{eqnarray}
Our choice of initial conditions then implies that all these coherences are identically zero. This means that the 
density operator $\bm{\rho}(t)$ can be written in the following block-diagonal matrix form:
\be\label{eq:densitym-blockdiagonal}
\bm{\rho}(t) = \left(\begin{array}{ccc}
              \rho_{00}(t) & 0 & 0 \\
              0 & \tilde{\bm{\rho}}(t) & 0 \\
              0 & 0 & \rho_{N+1,N+1}(t)
             \end{array}\right) ,
\ee
with $\tilde{\bm{\rho}}(t)$ the corresponding density matrix restricted to the subspace spanned by the nodes of the original network. 

Although we cannot make any detailed predictions for the dynamics of our system, we can make a general statement on the
temporal behavior of the population of the source node. From Eqs. \eqref{eq:lindbladplussource-drain} and
\eqref{eq:generator-sourcedrain} it follows that
\be
  \frac{d\rho_{00}(t)}{dt} = - \Gamma_k \,\rho_{00}(t),
\ee
leading to
\be\label{eq:gen-sol-pop-source}
  \rho_{00}(t) = e^{-\Gamma_k t}.
\ee
Thus, the population of the source will in any case decay exponentially. Since the source is incoherently coupled
to the network, eventually all its population will be transferred from the source to the network. 

\subsection{Transport efficiency}
In order to gain insight into the efficiency of the transport from the source
to the drain, we adopt the following definition of the transport efficiency which is similar to the one used
by Cao and Silbey \cite{Cao2009} and by Aspuru-Guzik et. al. \cite{Mohseni2008}: \emph{$\eta(\alpha)$ is the expected survival time (EST),
i.e. the expected amount of time that the excitation will remain inside the source and the network nodes}:
\begin{eqnarray}\label{eq:def-est}
\eta(\alpha) &=& \int_{0}^\infty dt  \sum_{k=0}^N \rho_{kk}(t,\alpha) \nonumber \\
          &=& \int_0^\infty dt \left[1-\rho_{N+1,N+1}(t,\alpha)\right] .
\end{eqnarray}
We stop to remark that the contribution of the source node to Eq. \eqref{eq:def-est} amounts
to a constant, since the decay of $\rho_{00}(t)$ is exponential, see Eq. \eqref{eq:gen-sol-pop-source}.
If $\eta(\alpha)$ is a small number this means that transport to the drain is relatively fast, and
vice-versa. $\eta(\alpha)$ can be calculated from the Laplace transforms of the populations $\rho_{kk}(t)$,
see appendix. \ref{app:eta-alpha-laplace}.
For instance, for a network consisting of only one node $\ket{1}$ that is connected to a
source and an drain, the EST $\eta(\alpha)$ can be easily
computed, see appendix \ref{app:monomer}:
\be
   \eta(\alpha) = 1 / \Gamma + 1 / \gamma ,
\ee
where $\Gamma$ and $\gamma$ are the source resp. drain strengths.
The EST is therefore independent of $\alpha$ and the transport becomes more efficient with
increasing values of $\gamma$ and $\Gamma$.

\section{Topologically disordered networks}
\begin{figure}
 \begin{center}
   \includegraphics[width=0.4\textwidth]{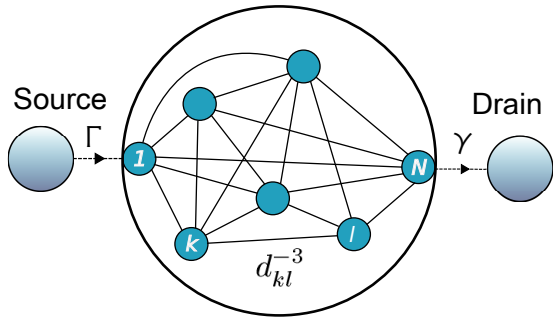}
 \end{center}
 \caption{Schematic representation of a single realisation of a topologically disordered network in
 a sphere with long-range dipole-dipole interactions. 
}\label{fig:sphere-network}
\end{figure} 

\begin{figure*}[!]
 \begin{center}
  \includegraphics[width=\textwidth]{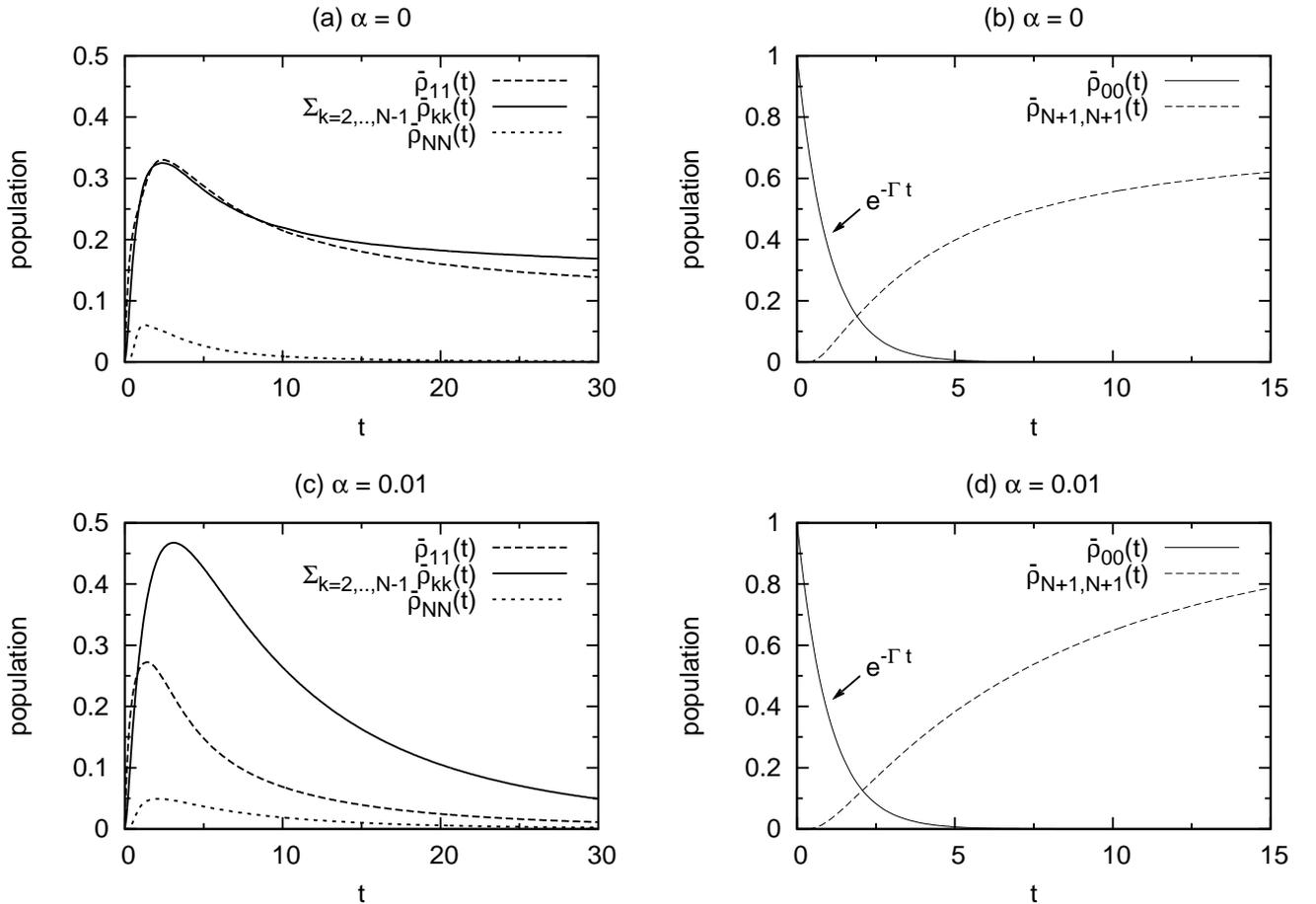}
 \end{center}
 \caption{ Populations of a topologically disordered network with $N=7$ nodes for $\alpha=0$, (a) and (b),
 and $\alpha=0.01$, (c) and (d), with $\Gamma = 0.5$ and $\gamma=1$. Figures (a) and (c) show the populations of 
 node $\ket{1}$ and node $\ket{N}$ as well as the sum of the populations of the remaining $N-2$ nodes of the network. 
 Figures (b) and (d) show the populations of the source and of the drain. 
 }\label{fig:pop-summed-network}
\end{figure*}

We now apply the above concepts to systems that exhibit
topological disorder and long-range dipole-dipole interactions. 
Here we model these systems by considering a random configuration of nodes in some bounded
region of space, say a sphere, which is coupled to an environment that induces diffusive
behaviour. To study the transport efficiency we furthermore connect a source and a
drain to the sphere. 
Previously, a similar system was studied, but without any coupling 
to the environment \cite{Mulken2010}. Also in \cite{Scholak2009,Scholak2011} a similar set-up was considered, but there the focus
was not on the quantum-to-classical crossover we study in this paper.

\subsection{The model}
We consider a network of $N$ sites located inside a sphere of radius $R$. At both
sides of the sphere we place a node and they are denoted by $\ket{1}$ resp. $\ket{N}$.
The other $N-2$ nodes are chosen randomly.
Furthermore, we connect a source $\ket{0}$ to $\ket{1}$ and a drain $\ket{N+1}$ to $\ket{N}$; their
respective coupling strengths are given by $\Gamma$ and $\gamma$, see Fig. \ref{fig:sphere-network} 
for an illustration of this system. As was noted before, we choose dipole-dipole interactions between the nodes, 
decaying as $d_{kl}^{-3}$ with $d_{kl}$ being the distance between the nodes
$k$ and $l$. The matrix elements of the Hamiltonian then take the following form:
\be
\bra{k} \mbf{H} \ket{l} = \left\{\begin{array}{cc}
                             - d_{kl}^{-3} & \mathrm{for}\,\, k \neq l \\
                             \sum_{j\neq k} d_{j k}^{-3} & \mathrm{for}\,\, k = l
                            \end{array}\right. .
\ee
In this particular case, the QSW equation descrbing the transition from the CTQW is given by
Eq. \eqref{eq:lindbladplussource-drain},
\be
\frac{d\bm{\rho}(t)}{dt} = (1-\alpha) \mathcal{L}_{\text{coh}} +  \alpha \sum_{k,l=1}^N | \mbf{H}_{kl} | \mathcal{D}(\mbf{L}_{kl}) + \mathcal{L}_{s+d} ,
\ee
with the source-drain superoperator
\be
\mathcal{L}_{s+d} (\bm{\rho}(t)) = \Gamma \,\mathcal{D}(\mbf{L}_{1,0},\bm{\rho}(t)) + \gamma \mathcal{D}(\mbf{L}_{N+1, N},\bm{\rho}(t)) .
\ee
For each realisation $r$ of the system, with $r=1,\ldots, \mathcal{R}$, we can numerically solve this 
equation and find the solution $\bm{\rho}^{(r)}(t)$. We calculate the ensemble average
\be
  \bar{\bm{\rho}}(t) = \frac{1}{\mathcal{R}} \sum_{r=1}^\mathcal{R} \bm{\rho}^{(r)}(t) ,
\ee
in order to obtain a global picture of the general dynamics of this system. 

\subsection{Numerical results}
\begin{figure}
 \begin{center}
   \includegraphics[width=\columnwidth]{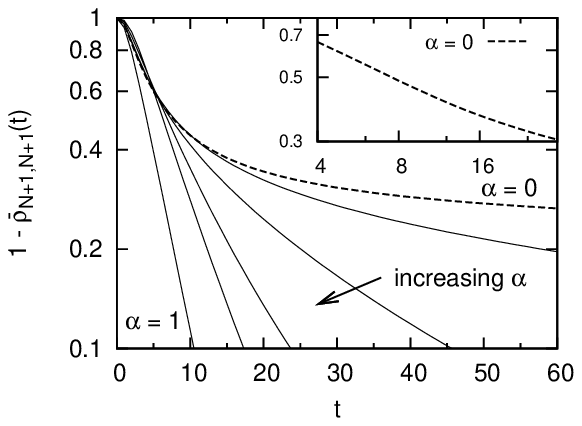}
 \end{center}
 \caption{Total population of the source-network nodes for various values of $\alpha=0,10^{-4},10^{-3},10^{-2},10^{-1},1$, $\Gamma=0.5$ and
  $\gamma=1$. The inset displays the curve for $\alpha=0$ in a log-log scale for the region that shows powerlaw decay.
 }\label{fig:pop-drain-varalpha}
\end{figure}

\begin{figure}
 \begin{center}
  \includegraphics[width=\columnwidth]{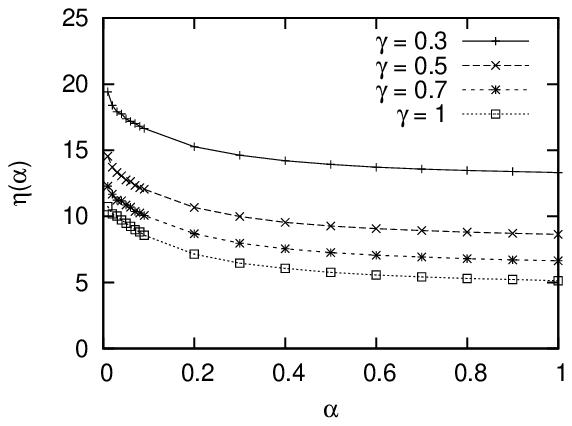}
 \end{center}
 \caption{The EST $\eta(\alpha)$ as a function of $\alpha$ for $\Gamma=0.5$ and for different values
 of $\gamma$. 
 }\label{fig:disordtop-efficiency}
\end{figure}
For all numerical results that will be discussed in this section, we assume that
$N=7$ for computational reasons. For the total number of realisations we take $\mathcal{R}=4000$. 
Without loss of generality we can take for the sphere unit radius, since other radii would merely 
rescale the average distances between the nodes and therefore would only provide a rescaling of the time axis. 
The precise value of the source strength $\Gamma$ is not particulary important since it only effects
the flow rate into the network, but does not qualitatively influence the flow to the drain:
its contribution to the EST $\eta(\alpha)$ is just a term equal to $1/\Gamma$. For the rest of this
section we therefore choose $\Gamma=0.5$.

As was shown in Eq. \eqref{eq:gen-sol-pop-source}, the initial excitation will decay exponentially
with rate $\Gamma$ from the source node into the connected node of the sphere, after which it flows
to the drain site. 
In Fig. \ref{fig:pop-summed-network} we show the populations of the system for the two different values of $\alpha=0$
and $\alpha=0.01$. Here we indeed observe an exponential decay of the population of the source (see $\rho_{00}(t)$ in
Figs. \ref{fig:pop-summed-network}(b) and \ref{fig:pop-summed-network}(d)), and an initially high population of node 
$\ket{1}$ (see $\rho_{11}(t)$ in Figs. \ref{fig:pop-summed-network}(a) and \ref{fig:pop-summed-network}(c)). 

An interesting observation arises from this figure: in the case of purely quantum mechanical transport
($\alpha=0$) we observe a form of short-to-intermediate time localization, with about 23\% of the population remaining in the network
after $t=40$, see Figs. \ref{fig:pop-summed-network}(a), \ref{fig:pop-summed-network}(b) and Fig. \ref{fig:pop-drain-varalpha} . 
Usually however, as time progresses there is still some population leaking into the drain. In the ensemble average eventually everything 
is then transferred to the drain. After switching on the environmental induced diffusion the localization effect completely vanishes, as can be
seen in Figs. \ref{fig:pop-summed-network}(c) and \ref{fig:pop-summed-network}(d). 

In Fig. \ref{fig:pop-drain-varalpha} we show 
the total population of the source-network system (excluding the drain) for different values of $\alpha$ (note the log-lin scale). 
For $\alpha = 0$ we observe, in the inset of Fig. \ref{fig:pop-drain-varalpha} that shows the curve in a log-log scale,
a power-law decay for intermediate times with $\eta(\alpha)\sim t^{-\beta}$ and $\beta\approx 0.21$,
which is characteristic for quantum walks in topologically disordered networks \cite{Mulken2010}. Already for small values
of $\alpha$, however, this behaviour quickly vanishes, ultimately leading to a pure exponential decay for $\alpha=1$. Here we
find that $\eta(\alpha) \sim e^{-\mu t}$ with $\mu = 0.247$. 
This means that classical diffusion already dominates over quantum transport for relatively small values of the 
environmental interaction. This conclusion holds true also for other values of $N$ (calculations which we do not show here).

Figure \ref{fig:disordtop-efficiency} shows a plot of the EST $\eta(\alpha)$ for this system for various
values of $\gamma$. As was already noted before, the transport efficiency is positively influenced by the 
environmental diffusion, since this removes the effects of localisation. We see that this is indeed also reflected in the computed 
transport efficiency, namely all curves for $\eta(\alpha)$ show a monotonous decay with increasing $\alpha$.
 We also observe that higher values of $\gamma$ lead to similar curves for the EST as a function
of $\alpha$, but with a lower overall value. Therefore larger trapping rates and larger values of $\alpha$ lead to faster transport to the drain.

This leads us to conjecture that for systems with quenched disorder and long-range (dipole-dipole) interactions transport will be on average 
dominated by classical diffusion, provided that the system is coupled to an environment that induces diffusive behavior. 

\section{Describing the disordered network with a dimer}
In this section we investigate
to what extent the topologically disordered network of the previous section can be described by 
an effective dimer model. 
To do this, we first focus on the properties of the EST $\eta(\alpha)$ for the general case of a heterodimer 
that is coupled to a source and drain. Its Hamiltonian is defined by:
\be\label{eq:hamiltdimer}
\mbf{H} = \left(\begin{array}{cc}
           0 & -V \\
           -V & \Delta \\
          \end{array}\right),
\ee
where $V$ represents the hopping rate between the nodes and $\Delta$ represents the energetic disorder between the two nodes.

In contrast to the case of the monomer (see Appendix \ref{app:monomer}), there exist multiple ways of connecting the source and drain to the dimer. Here
we focus only on the following case: we connect the source to node $\ket{1}$ and the drain to node $\ket{2}$. 
Only this configuration allows us to describe the topologically disordered network with a dimer. 

The master equation \eqref{eq:lindbladplussource-drain} can be solved analytically 
in this case. However, the full expression does not provide much insight, but leads to the exact expression for the EST
$\eta(\alpha)$, see Appendix \ref{app:efficiencydimer}. 
In the limit $\Delta\to 0$ we obtain:
\begin{equation}\label{eq:efficiency-dimer}
  \eta(\alpha) = \frac{2}{\gamma} + \frac{1}{\Gamma} + \frac{1}{V\alpha} - \frac{4(1-\alpha)^2}{\alpha V(4-8\alpha+6\alpha^2)+\alpha^2\gamma} .
\end{equation}
For $\gamma = \Gamma$ we find the following two limiting cases:
\begin{eqnarray}\label{eq:eff-diff-qm}
 \lim_{\alpha\to 0} \eta(\alpha ) &=& \frac{\gamma^2 + 4\Delta^2 + 3}{4V^2\gamma} \\
 \lim_{\alpha\to 1} \eta(\alpha ) &=& \frac{3}{\gamma} + \frac{1}{V}
\end{eqnarray}

\begin{figure}
 \begin{center}
  \includegraphics[width=\columnwidth]{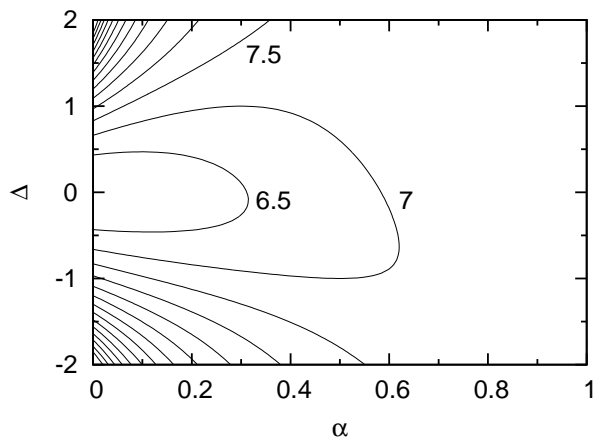}
 \end{center}
 \caption{Contour plot of the EST $\eta(\alpha)$ for a dimer as
 a function of $\alpha$ and the energy disorder $\Delta$ for $\gamma=\Gamma=0.5$ and $V=1$. The distance between the contour levels is $0.5$ 
 and the more outward lying contours have a higher value of $\eta(\alpha)$. We observe that $\eta(\alpha)$ is also positive for all values of $\Delta$.
 Furthermore, the EST approaches the constant value $1/V+3/\gamma=7$ in the limit $\alpha\to 1$.}\label{fig:eff-contourplot}
\end{figure}

For purely classical diffusion the transport efficiency is therefore completely determined by the hopping rate
$V$ and the source and drain rates $\gamma$.
This is however not the case for quantum mechanical
transport where $\Delta$ has a large influence on the transport efficiency. Larger values of $\Delta$ lead
to a quick increase in $\eta(\alpha)$, as can be observed from Fig. \ref{fig:eff-contourplot} and Eq. \eqref{eq:eff-diff-qm}.

\begin{figure}
 \begin{center}
  \includegraphics[width=\columnwidth]{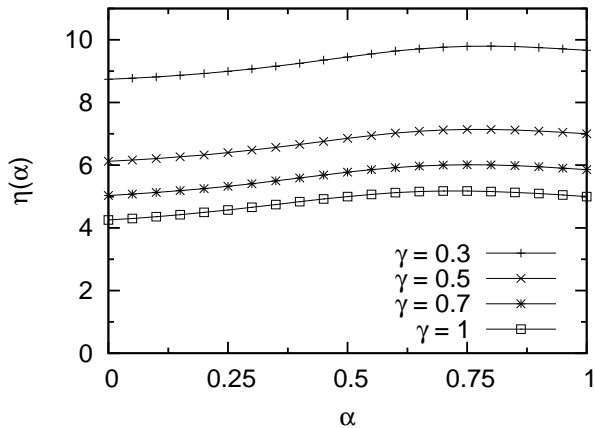}
 \end{center}
 \caption{The EST $\eta(\alpha)$ of a dimer for various values of $\gamma$, with $V=1$, $\Delta=0$
 and $\Gamma=0.5$. 
%  We observe that higher values of $\gamma$ lead to lower values
%  of the EST and therefore to faster transport. But unlike the disordered network, we don't observe a monotonous
%  decay with increasing $\alpha$, but instead a maximum around $\alpha=0.77$.
 }\label{fig:est-case2-nodisorder}
\end{figure}

\begin{figure}
 \begin{center}
  \includegraphics[width=\columnwidth]{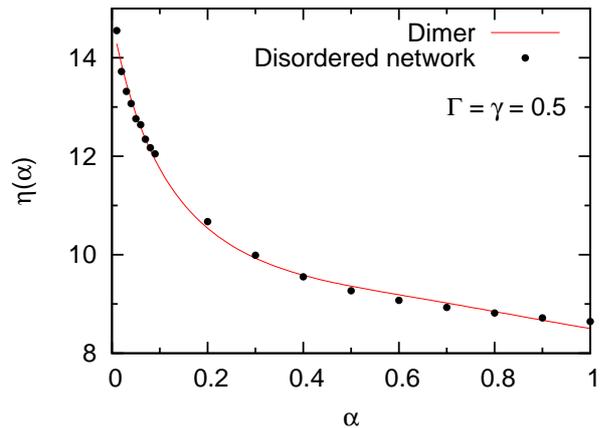}
 \end{center}
 \caption{Fit of the EST of the topologically disordered network with $\gamma = \Gamma = 0.5$ to that of a dimer. The fitting 
parameters are given by: $\gamma_d = 1.23$, $\Gamma_d = 0.19$, $V=0.61$ and $\Delta = 1.8$. Similar fits can be made for
 other values of $\gamma$ and $\Gamma$. One observes that the transport efficiency of a topologically disordered network can 
be well described with a dimer. }\label{fig:est-fit}
\end{figure}

In Fig. \ref{fig:est-case2-nodisorder} we show the EST $\eta(\alpha)$ for the case $\Delta = 0$ and for various values
of the drain strength $\gamma$. 
We observe that higher values of $\gamma$ lead to lower values
of the EST and therefore to faster transport. But unlike the disordered network, we don't observe a monotonous
decay with increasing $\alpha$, but instead a maximum around $\alpha=0.77$. 

The more general case of $\Delta \neq 0$
is shown as a contour plot in Fig. \ref{fig:eff-contourplot}. Here we observe that the EST is positive for all
values of $\Delta$ and that it approaches the constant value $1/V+3/\gamma$ in the limit $\alpha\to 1$. 
For values of $|\Delta | \gtrsim 1$ we do not observe a maximum of the EST anymore but instead a monotonous decay with increasing values of 
$\alpha$, resembling the EST for the topologically disordered network, see Fig. \ref{fig:disordtop-efficiency}.

After having studied the general properties of the EST of a dimer, we now proceed with fitting it to the
EST of the topologically disordered network.
In Fig. \ref{fig:est-fit} we provide a fit of the EST for the disordered network with $\Gamma = \gamma = 0.5$. The corresponding parameters
for the dimer are $\gamma_d = 1.23$, $\Gamma_d = 0.19$, $V=0.61$ and $\Delta = 1.8$, where the subscripts $d$ refer to 
the source and drain strengths of the dimer. In performing this fit, we have only fitted $\Gamma_d$, $\gamma_d$ and $V$ because
we have the freedom to choose the disorder strength $\Delta$, as long as $|\Delta|\gtrsim 1$. 
Higher values of $\Delta$ then correspond to larger values of both $\gamma_d$, $\Gamma_d$ and
$V$. This can be easily understood since in order to overcome a higher energy barrier one must increase the source strength and
at the same time also increase the drain strength. For other values of $\gamma$ and $\Gamma$ for the disordered network, the
fit behaves in a similar fashion. Therefore, the transport efficiency of a topologically disordered network can be well described
by modelling the system as a dimer with finite energy disorder. 

\section{Summary}
In this paper we studied the transport efficiency of an excitation moving from a source via a network to a drain. As a model
of many interesting physical systems such as ultra-cold Rydberg gases, we chose a topologically disordered network with long-range
interactions of dipole-dipole type. In particular, we focused on the crossover between purely quantum mechanical transport and
environmentally induced diffusion which we phenomenologically modeled by employing the recently introduced quantum stochastic walk.

Without any environmentally induced diffusion we observed localization at short to intermediate times of the excitation.
After switching on the environment this effect quickly vanished, with the total population in the network decaying exponentially
to the drain. This effect is furthermore largely indepedent on the number of nodes in the network. We can therefore
conclude that in a system with randomly placed nodes and dipole-dipole interactions between the nodes, transport is on average 
mainly dominated by classical diffusion, provided that the system is coupled to an environment that induces diffusive transport. 

As a measure for the transport efficiency we used the expected survival time (EST) $\eta(\alpha)$, defined to be the expected amount
of time that it takes to transfer the excitation from the source to the drain. We found that for all values of the source
and drain strengths, transport is most efficient in the purely classical case. Furthermore, we showed that on the
level of the EST the topologically disordered system can be mapped on a dimer with a finite energy difference between the nodes. 

\acknowledgments{
 We gratefully acknowledge support from the Deutsche Forschungsgemeinschaft (DFG) and the Fonds der Chemischen Industrie.
 Furthermore, we thank A. Anishchenko and L. Lenz for useful discussions.  
}

\appendix

\section{Random walks in the Lindblad formalism}\label{app:randomwalk}
Here we provide a short derivation on how to obtain a random walk in the Lindblad formalism.
In order to provide this connection, we consider Lindblad operators of the form $\mbf{L}_{kl} = \ket{k}\bra{l}$ for
$k,l=1,\ldots N$. Since the Lindblad operators form a complete orthonormal basis, we can expand the dissipators
$\mathcal{D}(\mbf{L}_{kl},\bm{\rho})$ in these operators:
\be\label{eq:explicit-dissipators}
   \mathcal{D}(\mbf{L}_{kl}) = \rho_{ll} \mbf{L}_{kk} - \rho_{ll} \mbf{L}_{ll} - 
   \frac{1}{2}\sum_{j=1,j\neq l}^N \left(\rho_{lj} \mbf{L}_{lj} + \rho_{jl} \mbf{L}_{jl} \right) .
\ee
Suppose now that we reparametrize the Lindblad equation as in Eq. \eqref{eq:qsw-alpha} and take $\alpha=1$. The
dynamics is then completely governed by the dissipator $\mathcal{D}(\bm{\rho})$. Inserting the above 
expansion in the Lindblad equation results in the following differential equations for the diagonal
components of the density matrix:
\be\label{eq:ctrw-from-diss}
   \frac{d\rho_{kk}}{dt} = \sum_{l=1}^{N}( \lambda_{kl} \rho_{ll} - \lambda_{lk}\rho_{k k} ) ,
\ee
and similarly for the offdiagonal components:
\be
\frac{d\rho_{km}}{dt} = -\frac{1}{2} \sum_{j=1}^{N} (\lambda_{j k} + \lambda_{j m} ) \rho_{km} .
\ee
This implies that the equations for the coherences $\rho_{km}$ decouple from the populations $\rho_{kk}$. 
Furthermore, we can rewrite Eq. \eqref{eq:ctrw-from-diss} for the populations in the form
of Eq. \eqref{eq:mastereq-ctrw}, with
\be
  T_{kl} = \left(1-\delta_{k,l}\right) \lambda_{k l} + \delta_{k,l} \sum_{j=1,j\neq k}^N \lambda_{j,k} . 
\ee
The matrix $\mbf{T}$ formed by the coefficients $T_{kl}$ can then be interpreted as the transfer matrix
for a CTRW. The coupling constants $\lambda_{kl}$ therefore represent the transfer rates of moving
from node $\ket{l}$ to node $\ket{k}$. 

\section{Relation to modelling traps with effective Hamiltonians}\label{app:effectiveHamilt-trap}
We now provide a short proof of the equivalence of modelling a drain with Lindblad operators and of modelling a
drain by introducing an imaginary part to the Hamiltonian, as was done in previous works \cite{Mulken2007,Agliari2010}. Suppose 
that we connect a drain to node $\ket{N}$ with strength $\gamma_N$. One can then introduce the effective Hamiltonian
$\mbf{H}_{eff} = \mbf{H} - i \hat{\mbf{\Gamma}}$ with $\hat{\mbf{\Gamma}} = \gamma_N \ket{N}\bra{N}$. This leads to the following modification
to the Liouville equation:
\be
   \frac{d\bm{\rho}_S}{dt} = -i \left[ \mbf{H}, \bm{\rho}_S\right] - \left\{\hat{\mbf{\Gamma}}, \bm{\rho}_S \right\} .
\ee
Expanding the last term of this equation in terms of the Lindblad operators $\mbf{L}_{kl}$ results in
\be
\frac{1}{2\gamma_N}\left\{\hat{\mbf{\Gamma}}, \bm{\rho} \right\} = \rho_{NN} \mbf{L}_{NN} + \frac{1}{2}\sum_{j=1}^{N-1} \left(\rho_{N j} \mbf{L}_{Nj} + \rho_{j N}\mbf{L}_{j N} \right) .
\ee
These are clearly the same terms as the one that would be obtained from Eqns. \eqref{eq:explicit-dissipators}
in combination with Eq. \eqref{eq:generator-sourcedrain}, after projecting the dissipator onto the subspace
spanned by the network nodes. The only difference is the factor of two that arises in the coupling strength $\gamma_N$,
but this is a matter of convention. 

\section{Computation of $\eta(\alpha)$ with a Laplace transform of the density matrix}\label{app:eta-alpha-laplace}
The EST $\eta(\alpha)$ can also be written in terms of the Laplace transforms $\hat{\rho}_{kk}(s)$
of the diagonal components of the density matrix:
\begin{eqnarray}
  \eta(\alpha) &=& \lim_{s\to 0} \sum_{k=0}^N \int_0^\infty dt  e^{-st}  \rho_{kk}(t) \\
            &=& \lim_{s\to 0} \sum_{k=0}^N \hat{\rho}_{kk}(s) .
\end{eqnarray}
Writing the components of the density matrix in a vector $\vec{\rho}(t)=(\rho_{00}(t),\rho_{11}(t),\ldots, \rho_{N+1,N+1}(t) )$
leads to 
\be
  \frac{d\vec{\rho}(t)}{dt} = \mathcal{L} \vec{\rho}(t) ,
\ee
with $\mathcal{L}$ being the superoperator including all coherent, diffusive and incoherent terms. 
This then results in the following equation for their Laplace transforms:
\be
  \left(s\, \mathbb{I} - \mathcal{L} \right) \hat{\vec{\rho}}(s) = \vec{\rho}(0) .
\ee

\section{Complete expressions for the EST of the dimer}\label{app:efficiencydimer}
For the dimer with a nonzero value of the energy offset $\Delta$ we find the following expression for EST $\eta(\alpha)$:
\be
  \eta(\alpha) = \frac{2}{\gamma} + \frac{1}{\Gamma} + \frac{1}{\alpha}\left[ \frac{1}{V} - \frac{f(\alpha)}{g(\alpha)} \right]
\ee
The function $f(\alpha)$ is given by:
\be
   f(\alpha) = 4(1-\alpha^2)(2 V\alpha + \gamma + \alpha\Delta), 
\ee
and the function $g(\alpha)$ is given by:
\begin{eqnarray}
 g(\alpha) &=& 4V^2\alpha(2+\alpha(3\alpha-4)) + 4V(1+2(\alpha-1)\alpha)(\gamma + \alpha\Delta) \nonumber \\
           && + \alpha(\gamma^2+2\alpha\gamma\Delta + (4+\alpha(5\alpha-8))\Delta^2) 
\end{eqnarray}
Note that both $f(\alpha)$ and $g(\alpha)$ are independent of the source strength $\Gamma$.

\section{A monomer with source and drain}\label{app:monomer}

Here we provide a simple example to illustrate the concepts introduced in section \ref{sec:sourcedrain}.
We take the network to be only a single node $\ket{1}$ such that
the source-network-drain system consists of three nodes in total. For notational simplicity we will write 
$\Gamma = \Gamma_1$ and $\gamma = \gamma_1$. The Hamiltonian $\mbf{H}$ is proportional to $\ket{1}\bra{1}$. According
to Eqns. \eqref{eq:lindbladplussource-drain}-\eqref{eq:vanishing-coherences-2}, 
we obtain a master equation where all non-diagonal elements are decoupled, each
of which acquires a simple exponential solution. Our choice of the initial condition $\bm{\rho}(0) = \ket{0}\bra{0}$
thus yields $\rho_{kj}(t) = 0$ for $k\neq j$. For the diagonal elements we find the following equations:
\begin{eqnarray}
 \dot{\rho}_{00}(t) &=& -\Gamma \rho_{00}(t) \\
 \dot{\rho}_{11}(t) &=& \Gamma \rho_{00}(t) - \gamma \rho_{11}(t) \\
 \dot{\rho}_{22}(t) &=& \gamma \rho_{11}(t) .
\end{eqnarray}
which have the solutions 
\begin{eqnarray}
\rho_{00}(t) &=& \exp{\left(-\Gamma t\right)} \\
\rho_{11}(t) &=& \frac{\Gamma}{\Gamma - \gamma} \left( \exp{(-\gamma t)} - \exp{(-\Gamma t)}\right) \\
\rho_{22}(t) &=& 1 - \frac{ \Gamma \exp(-\gamma t) - \gamma \exp(-\Gamma t)}{\Gamma -\gamma} ,
\end{eqnarray}
The dynamics of a monomer is thus independent of the value of $\alpha$. 

One also easily sees that the EST $\eta(\alpha)$ is given by:
\be
   \eta(\alpha) = \frac{1}{\Gamma} + \frac{1}{\gamma}
\ee
Therefore, the transport efficiency always increases with increasing $\gamma$ and
$\Gamma$. For example a high input with high output leads to a high efficiency
and vice-versa. 

% \bibliographystyle{h-physrev3.bst} 
% \bibliography{/home/piet/Documents/journal-papers/jabref-database.bib}

\begin{thebibliography}{10}

\bibitem{Anderson1998}
W.~Anderson, J.~Veale, and T.~Gallagher,
\newblock \prl {\bf 80}, 249 (1998).

\bibitem{Westermann2006}
S.~Westermann {\em et~al.},
\newblock Eur. Phys. J. D {\bf 40}, 37 (2006).

\bibitem{Saffman2010}
M.~Saffman, T.~Walker, and K.~Molmer,
\newblock Rev. Mod. Phys. {\bf 82}, 2313 (2010).

\bibitem{Jaksch2000}
D.~Jaksch {\em et~al.},
\newblock \prl {\bf 85}, 2208 (2000).

\bibitem{Cote2006}
R.~C\^ot\'e, A.~Russell, E.~Eyler, and P.~Gould,
\newblock New. J. Phys. {\bf 8}, 156 (2006).

\bibitem{Kenkre1982}
V.~Kenkre and P.~Reineker,
\newblock {\em Exciton Dynamics in Molecular Crystals and Aggregates}
  (Springer, Berlin, 1982).

\bibitem{Reineker2004}
P.~Reineker,
\newblock J. Lumin. {\bf 108}, 149 (2004).

\bibitem{Collini2010}
E.~Collini {\em et~al.},
\newblock \nat {\bf 463}, 644 (2010).

\bibitem{Fleming2004}
G.~Fleming and G.~Scholes,
\newblock \nat {\bf 431}, 256 (2004).

\bibitem{Cheng2006}
Y.~Cheng and R.~Silbey,
\newblock \prl {\bf 96}, 028103 (2006).

\bibitem{Engel2007}
G.~Engel {\em et~al.},
\newblock \nat {\bf 446}, 782 (2007).

\bibitem{Panitchayangkoon2010}
G.~Panitchayangkoon {\em et~al.},
\newblock P. Natl. Acad. Sci. USA {\bf 107}, 12766 (2010), 1001.5108.

\bibitem{Vaziri2010}
A.~Vaziri and M.~Plenio,
\newblock New. J. Phys. {\bf 12}, 085001 (2010).

\bibitem{Cao2009}
J.~Cao and R.~Silbey,
\newblock J. Phys. Chem. A {\bf 113}, 13825 (2009).

\bibitem{Caruso2009}
F.~Caruso, A.~Chin, A.~Datta, S.~Huelga, and M.~Plenio,
\newblock \jcp {\bf 131}, 105106 (2009).

\bibitem{Rebentrost2009}
P.~Rebentrost, M.~Mohseni, I.~Kassal, S.~Lloyd, and A.~Aspuru-Guzik,
\newblock New. J. Phys. {\bf 11}, 033003 (2009).

\bibitem{Mohseni2008}
M.~Mohseni, P.~Rebentrost, S.~Lloyd, and A.~Aspuru-Guzik,
\newblock \jcp {\bf 129}, 174106 (2008).

\bibitem{Diehl2008}
S.~Diehl {\em et~al.},
\newblock Nat. Phys. {\bf 4}, 878 (2008).

\bibitem{Verstraete2009}
F.~Verstraete, M.~Wol, and J.~Ignacio~Cirac,
\newblock Nat. Phys. {\bf 5}, 633 (2009).

\bibitem{Farhi1998}
E.~Farhi and S.~Gutmann,
\newblock \pra {\bf 58}, 915 (1998).

\bibitem{Muelken2011}
O.~M\"ulken and A.~Blumen,
\newblock Phys. Rep. {\bf 502}, 37 (2011).

\bibitem{VanKampen1990}
N.~V. Kampen,
\newblock {\em Stochastic Processes in Physics and Chemistry} (North Holland,
  Amsterdam, 1990).

\bibitem{Whitfield2010}
J.~Whitfield, C.~A. Rodriguez-Rosario, and A.~Aspuru-Guzik,
\newblock \pra {\bf 81}, 022323 (2010).

\bibitem{HakenStroblBook}
H.~Haken and G.~Strobl,
\newblock {\em The Triplet State} (Cambridge University Press, 1967).

\bibitem{Haken1972}
H.~{Haken} and P.~{Reineker},
\newblock Z. Phys. {\bf 249}, 253 (1972).

\bibitem{Michaelis2006}
B.~Michaelis, C.~Emary, and C.~Beenakker,
\newblock Europhys. Lett. {\bf 73}, 677 (2006).

\bibitem{Groth2006}
C.~Groth, B.~Michaelis, and C.~Beenakker,
\newblock \prb {\bf 74}, 125315 (2006).

\bibitem{Sarovar2011}
M.~Sarovar, Y.-C. Cheng, and K.~Whaley,
\newblock \pre {\bf 83}, 011906 (2011).

\bibitem{BreuerOpenQS}
H.~Breuer and F.~Petruccione,
\newblock {\em The theory of open quantum systems} (Oxford University Press,
  2010).

\bibitem{Lindblad1976}
G.~Lindblad,
\newblock Commun. Math. Phys. {\bf 48}, 119 (1976).

\bibitem{SchijvenUnpublished}
P.~Schijven, A.~Blumen, and O.~M\"ulken,
\newblock \emph{In progress} .

\bibitem{Agliari2010}
E.~Agliari, O.~M\"ulken, and A.~Blumen,
\newblock Int. J. Bifurcat. Chaos {\bf 20}, 271 (2010).

\bibitem{Mulken2007}
O.~M\"ulken {\em et~al.},
\newblock \prl {\bf 99}, 090601 (2007).

\bibitem{Mulken2010}
O.~M\"ulken and A.~Blumen,
\newblock Phys. E {\bf 42}, 576 (2010).

\bibitem{Scholak2009}
T.~Scholak, F.~de~Melo, T.~Wellens, F.~Mintert, and A.~Buchleitner,
\newblock \pre {\bf 83}, 021912 (2009).

\bibitem{Scholak2011}
T.~Scholak, T.~Wellens, and A.~Buchleitner,
\newblock (2011), 1103.2944.

\end{thebibliography}

\end{document}